\documentclass[fleqn,usenatbib]{mnras}

\usepackage[T1]{fontenc}

\newcommand{\ith}{\ensuremath{^{\rm th}}}

\newcommand{\alphafe}{[$\alpha$/Fe]}

\newcommand{\ha}{high-$\alpha$}

\newcommand{\lowalpha}{low-$\alpha$~}

\newcommand{\rbir}{R$_b$}

\newcommand{\feh}{[Fe/H]}

\DeclareRobustCommand{\VAN}[3]{#2}
\let\VANthebibliography\thebibliography
\def\thebibliography{\DeclareRobustCommand{\VAN}[3]{##3}\VANthebibliography}

\usepackage{graphicx}	
\usepackage{amsmath}	
\usepackage{amssymb}	
\usepackage{newtxtext,newtxmath}
\usepackage{lineno}

\title[Finding Stellar Birth Radii]{There is No Place Like Home — Finding Birth Radii of Stars in the Milky Way}

\author[Lu et al.]{
Yuxi (Lucy) Lu$^{1, 2, 3}$\thanks{E-mail: lucylulu12311@gmail.com},
Ivan Minchev$^{3}$,
Tobias Buck$^{4,5}$,
Sergey Khoperskov$^{3}$,
Matthias Steinmetz$^{3}$,
\newauthor
Noam Libeskind$^{3}$,
Gabriele Cescutti$^{6,7}$,
Ken C. Freeman$^{8}$,
Bridget Ratcliffe$^{3}$
\\
$^{1}$Department of Astronomy, Columbia University, 550 West 120\ith\ Street, New York, NY, USA\\
$^{2}$American Museum of Natural History, Central Park West, Manhattan, NY, USA\\
$^{3}$Leibniz Institute for Astrophysics Potsdam, An der Sternwarte 16, 14482 Potsdam, Germany\\
$^{4}$Universit\"at Heidelberg, Interdisziplin\"ares Zentrum f\"ur Wissenschaftliches Rechnen, Im Neuenheimer Feld 205, Heidelberg, Germany\\
$^{5}$Universit\"at Heidelberg, Zentrum f\"ur Astronomie, Institut f\"ur Theoretische Astrophysik, Albert-Ueberle-Straße 2, Heidelberg, Germany\\
$^{6}$INAF, Osservatorio Astronomico di Trieste, via G.B. Tiepolo 11, Trieste, Italy\\
$^{7}$Dipartimento di Fisica, Sezione di Astronomia, Universit. di Trieste, via G.B. Tiepolo 11, Trieste, Italy\\
$^{8}$Research School of Astronomy \& Astrophysics, Australian National University, ACT 2611, Australia
}

\date{Accepted XXX. Received YYY; in original form ZZZ}

\pubyear{2021}

\begin{document}
\label{firstpage}
\pagerange{\pageref{firstpage}--\pageref{lastpage}}
\maketitle

\begin{abstract}
Stars move away from their birthplaces over time via a process known as radial migration, which blurs chemo-kinematic relations used for reconstructing the Milky Way (MW) formation history. 
To understand the true time evolution of the MW, one needs to take into account the effects of this process. We show that stellar birth radii can be derived directly from the data with minimum prior assumptions on the Galactic enrichment history. This is done by first recovering the time evolution of the stellar birth metallicity gradient, $d\mathrm{[Fe/H]}(R, \tau)/dR$, through its inverse relation to the metallicity range as a function of age today, allowing us to place any star with age and metallicity measurements back to its birthplace, \rbir. Applying our method to a large, high-precision data set of MW disk subgiant stars, we find a steepening of the birth metallicity gradient from 11 to 8 Gyr ago, which coincides with the time of the last massive merger, Gaia-Sausage-Enceladus (GSE).
This transition appears to play a major role in shaping both the age-metallicity relation and the bimodality in the [$\alpha$/Fe]-[Fe/H] plane. By dissecting the disk into mono-\rbir\ populations, clumps in the low-\alphafe\ sequence appear, which are not seen in the total sample and coincide in time with known star-formation bursts, possibly associated with the Sagittarius Dwarf Galaxy. We estimated that the Sun was born at $4.5\pm 0.4$~kpc from the Galactic center. Our \rbir\ estimates provide the missing piece needed to recover the Milky Way formation history.
\end{abstract}




\section{Introduction} \label{sec:intro}
Galactic Archaeology aims to understand the formation history of the Milky Way (MW) either via observing high-redshift galaxies or by inferring the history from the current-day MW data. While observing galaxies at different lookback times can help us understand the physical processes that govern the formation of galaxies, the unique and detailed formation history of the MW can only be inferred from itself. 

In order to understand the MW formation history from just the present day data, it is crucial to have accurate and precise measurements of abundances and ages of stars in the Galaxy. Stellar abundances can act as fossils as it is believed that most of the element composition of a star does not change much over its lifetime. If this is true, by combining stellar ages and abundances, we can gain insight into the formation and evolution of the Galaxy \citep[e.g.][]{Freeman2002, Ratcliffe2020, Horta2022, Lu2022c}, as well as the nucleosynthetic channels of chemical enrichments \citep[e.g.][]{Ting2012, Weinberg2019, Griffith2021,Ratcliffe23a}.

Recent large spectroscopic surveys, such as the Apache Point Observatory Galactic Evolution Experiment (APOGEE) \citep{Majewski2017}, Large Sky Area Multi-Object Fibre Spectroscopic Telescope (LAMOST) \citep{LAMOST}, GALactic Archaeology with HERMES (GALAH) \citep{Silva2015, Buder2019}, and Gaia-ESO \citep{gaiaeso} have provided an enormous amount of spectra for stars in our Galaxy.
Detailed element abundances have been derived form these spectra and contributed greatly to the field of Galactic Archaeology. 
However, due to the lack of understanding of the interstellar medium, Earth's atmosphere, instrumental noise, and the star itself, it is hard to achieve abundance precision less than $\sim0.01$ dex for large samples \citep[e.g.][]{Asplund2005,Asplund2009}. Even more, \cite{Anguiano2018} pointed out that metallicities derived from APOGEE and LAMOST disagree on the order of 0.1 dex. As a result, many data-driven approaches have been used to improve the abundance measurements and show promising results \citep[e.g.][]{Bedell2014, Ness2015, Xiang2019}. On the other hand, the age of a star is not a direct observable but an estimation of its evolutionary stage. This requires understanding of its complex structure and thus, most age-dating methods have uncertainties above $\sim20\%$ \citep[for a detailed review on stellar ages, see][]{Soderblom2010}. However, within these methods, isochrone fitting for main-sequence turn-off stars (MSTO) or sub-giants is able to provide accurate ages for stars with precise and accurate abundance and photometric measurements. For example, \cite{Xiang2022} were able to measure ages for a large sample of sub-giant stars in LAMOST with a median uncertainty of only 7.5\%. 

Unfortunately, even with precise and accurate age and abundance measurements, inferring the MW formation history from only the current day data is still difficult, as stars have moved away from their birth location overtime via radial migration.
During this process, angular momenta of stars are permanently changed due to their interactions with resonances caused by the spiral arms \citep[e.g.][]{Sellwood2002,Roskar2008} and the central bar \citep[e.g.][]{Minchev2010, dimatteo13, Khoperskov2020}.
Stars that have migrated cannot be distinguished from local ones by their kinematics alone, and thus the process has been identified from the scatter in the age-metallicity relation in local stars and the increasing scatter with age in the radial gradient of mono-age populations (e.g., \citealt{Anders2017}). This process has been shown to flatten the intrinsic radial abundance gradient, or the birth gradient, in the disk significantly over time \citep[][]{Minchev2012, Minchev2013, Vincenzo2020} and erases formation signatures, especially for older stars. 
This means that trends in stellar age is does not reflect the true evolution with lookback time; thus inferring the MW formation history from mono-age population without taking into account radial migration is likely to provide misleading conclusions \citep[e.g.,][]{Anders2017, Ratcliffe23b}.

The most straightforward approach to take into account radial migration would be to estimate the stellar birth radii, \rbir. \cite{Minchev2018} presented a largely model-independent approach for estimating \rbir\ based only on precise metallicity and age estimates, which was applied to the local AMBRE:HARPS \citep{delaverny13} and HARPS-GTO \citep{Adibekyan2012} samples. This technique relied on the following assumptions: (1) the gas is well mixed azimuthally, (2) stars are born with a narrow metallicity range at a certain radius at any given time, (3) the MW disk formed inside-out, and (4) the ISM [Fe/H]($\tau$, R) evolved smoothly with both Galactic disk radius, R, and cosmic time, $\tau$. One way to test whether these assumptions are true is to analyze simulations. 

Unlike observations, simulations provide the exact formation history with no measurement uncertainties. Many simulations have successfully reproduced key observations in MW-like disk galaxies \citep[e.g.,][]{Aumer2013,Stinson2013,martig14a,Marinacci2014,Wang2015,Grand2017,Hopkins2018,Buck2019c,Buck2020b}. Although feedback mechanisms and parameter selections can greatly affect the results \citep[e.g.,][]{Keller2019,Dutton2019,Munshi2019,Blancato2019,Buck2020b}, simulations are able to provide important insights on the formation history of the MW. For example, in \cite{Lu2022}, we tested the assumptions in \cite{Minchev2018} using the NIHAO-UHD simulations \citep{Buck2018} and found that assumptions (1), (2), and (3) are indeed satisfied once a rotationally supported stellar disk has started to form, while (4) can be violated during gas-rich mergers. More specifically, during such events, metal-poor gas from a merging satellite can quickly dilute the gas in the host disk and cause the overall metallicity in the outskirts to decrease. This would then result in steepening of the metallicity gradient disrupting its monotonic evolution \citep[for a detail analysis on how this can happen, see][]{Buck2023}. Naturally, taking into account the effects of satellite infall in estimating \rbir\ is the obvious next step, considering that the MW likely experienced a massive merger event (the Gaia-Sausage-Enceladus, hereafter GSE) $\sim$ 8-10 Gyr ago \citep[e.g.,][]{Belokurov2018, Helmi2018, Gallart2019, Grunblatt2021, Borre2022, Buck2023}.

We introduce here a new empirical method for recovering both the evolution of the MW disk metallicity with radius and time, $d\mathrm{[Fe/H]}(R, \tau)/dR$, and the birth radius, \rbir, of stars, simply based on their age and metallicity measurements. The observational and simulation data used are described in Sec.~\ref{subsection:data}. The method to derive the metallicity time evolution is presented in Sec.~\ref{subsection:method}. Discussion on the formation of the MW age-metallicity relation (AMR) and of the \alphafe\ bimodality can be found in (section~\ref{sec:results}). Sec.~\ref{subsec:limiation} points out the limitation of this work and we conclude in Sec.~\ref{subsec:future}.

\section{Data}\label{subsection:data}
\subsection{Observational data}

In order to infer the history of the MW, both accurate and precise abundances and ages are needed. We use the largest and most precise sample of isochrone stellar ages to-date (median age uncertainty of 7.5\%) provided by \cite{Xiang2022}. These authors estimated ages for $\sim$ 250,000 subgiant stars from the LAMOST survey, covering an extended Galactic disk area (about $6<R<12$~kpc, with most of the stars between $7<R<10$~kpc), with isochrone fitting.

For this study, we selected stars satisfying the following criteria:
\begin{itemize}
    \item \feh\ $>$ -1 \& eccentricity $<$ 0.5 \& $\lvert z\rvert<1$~kpc, to select disk stars.
    \item age uncertainty $<$ 1 Gyr, to select stars with age error smaller than the width of the age bin that will be used to infer the metallicity evolution.
    \item \feh\ uncertainty $<$ 0.05 dex, to select stars with good metallicity measurements.
    \item age~$<$~13~Gyr, as a rotationally supported MW disk most likely did not exist before that time \citep{Belokurov2022, Rix2022, Conroy2022}. 
\end{itemize}

After applying these cuts, we are left with 77,475 subgiant stars with an average age and metallicity uncertainty of 0.32 Gyr and 0.03 dex, respectively. 

\begin{figure*}
    \centering
    \includegraphics[width=0.8\textwidth]{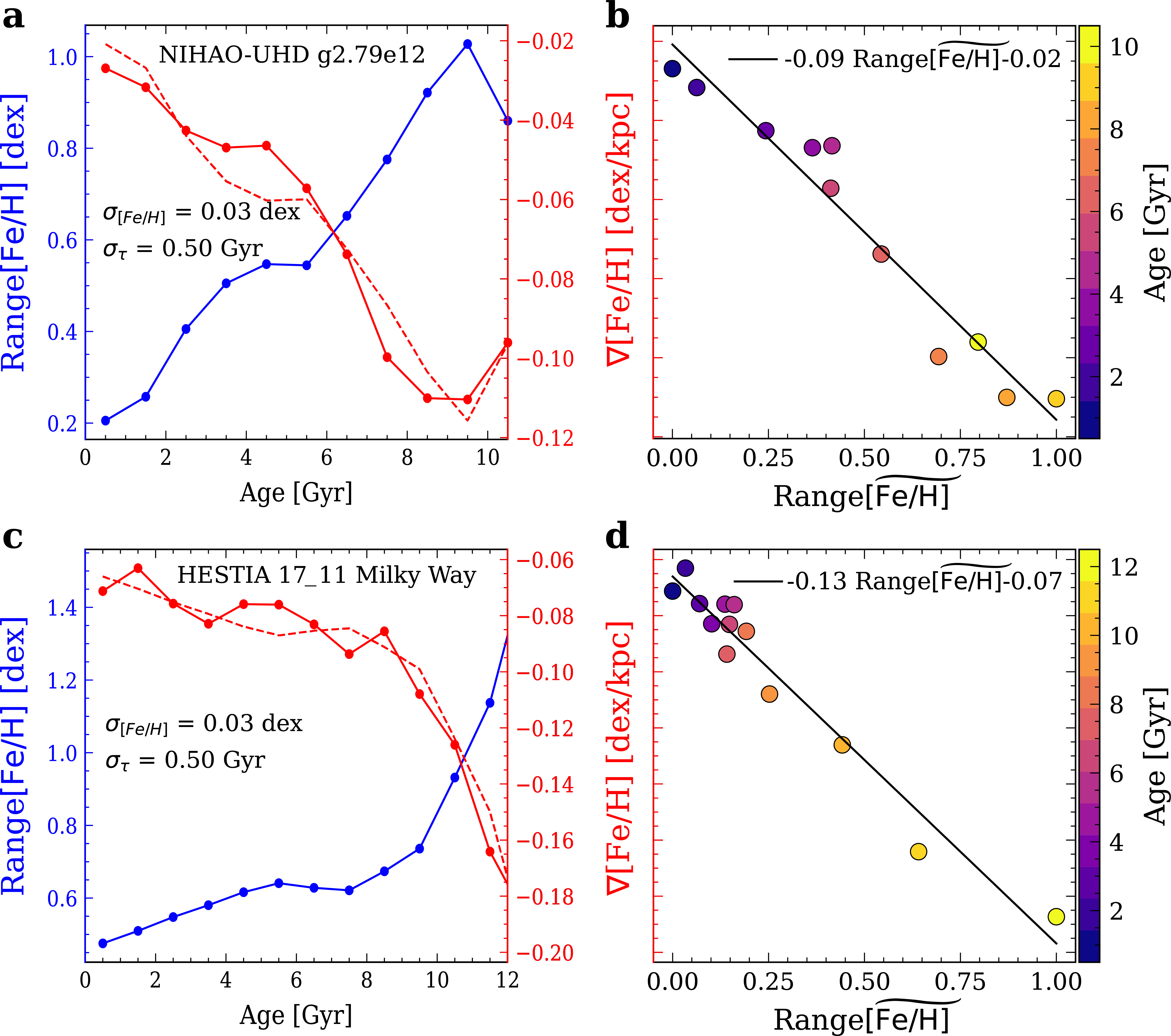}
    \caption{Relation between Range[Fe/H](age) and $\mathrm{\nabla [Fe/H](\tau)}$ in two cosmological MW-like simulations. The sample selection from both simulations matches that of our LAMOST sample, including uncertainties in [Fe/H] and age, as indicated in the left panels. {\bf Panel a}: The Range[Fe/H](age) (95\%-5\%-tiles in metallicity; blue line) and the birth gradient, $\mathrm{\nabla [Fe/H](\tau)}$ (solid red line) estimated for the NIHAO-UHD g2.79e12 galaxy \citep{Buck2020} show a well-defined anti-correlation, which is linear in nature (PCC $=-0.97$), as shown by the black line fit in {\bf panel b}. Inferring $\mathrm{\nabla [Fe/H](\tau)}$ directly from the line fit results in an error of about 9\%, on average, of the true birth metallicity gradient (see red dashed line in panel a). Similarly to the data (Fig.~\ref{fig:fehevo}a), a steepening in the gradient is seen early on, associated with a massive merger prior to that. This lends credibility to our interpretation that the gradient fluctuation we find in the data is related to the GSE merger. 
    {\bf Panels c,d:} Same as panels a,b but for the HESTIA 17\_11 MW analog \citep{Libeskind2020,Khoperskov2022}. The inferred birth $\mathrm{\nabla [Fe/H](\tau)}$ is, on average, within 7\% of the true one, and the line fit has a PCC $=-0.98$. 
    }
    \label{fig:1}
\end{figure*}

\subsection{Simulation data}
\label{sec:sims}

The simulated galaxies studied in this work are from the NIHAO-UHD \citep{Buck2018, Buck2020b} and HESTIA \citep{Libeskind2020, Khoperskov2022} projects, both of which present high-resolution cosmological hydrodynamical simulations of MW-mass galaxies.

The NIHAO-UHD galaxy model was calculated using a modified version of the smoothed particle hydrodynamics (SPH) solver GASOLINE2 \citep{Wadsley2017} and the simulation is calculated from cosmological initial conditions and star formation and feedback are modelled following the prescriptions in \citep{Stinson2006,Stinson2013}. 
Details about these simulations can be found in Table 1 by \cite{Buck2020b}. 

The HESTIA simulations are M31/MW pairs produced using the code \texttt{AREPO} \citep{Springel2010, Pakmor2016, Weinberger2020} and the galaxy formation model from \texttt{AURIGA} \citep{Grand2017}. Details of this simulation suite can be found in \cite{Libeskind2020}.

For this study, we focus on the g2.79e12 simulation from NIHAO-UHD and the 17\_11 M31/MW pair from HESTIA, as they are the highest resolution simulations in each simulation suite, both are able to reproduce MW-like properties, and have undergone MW-like merger histories \citep{Buck2020, Libeskind2020, Khoperskov2022}. To mimic the data, we selected only disk stars ([Fe/H] $>$ -1, eccentricity $<$ 0.5 kpc, |z| $<$ 1 kpc) with radii between 7-10 kpc at redshift zero. We have chosen a radial range a bit smaller than the data ($\sim6-12$ kpc), because only a small number of stars are found outside that.

\section{Methods: Reconstruction of the metallicity enrichment history}\label{subsection:method}

To reconstruct the MW metallicity evolution with radius and time, we make the following sensible assumptions: (1) the gas is well mixed in Galactic azimuth as observed in the MW \citep{Arellano-Cordova21,Esteban2022} and external galaxies \citep{Kreckel2019}, as well as seen in cosmological simulations throughout the disk formation \citep{Lu2022}, (2) the MW disk formed from the inside out, which is now well established \citep{matteucci89, Sharma2021}, and (3) there is a well-defined linear relation between metallicity and \rbir, as seen in simulations of galactic disk formation \citep{Lu2022}.

While our method is similar to that used in a recent work \citep{Minchev2018}, the improvement here, besides the much larger data set with excellent age estimates, is that the time evolution of the metallicity slope is recovered from the AMR, rather than from the distribution of birth radii of mono-age populations, as described below.

In the following Sec. we test our method on the simulations, and in Sec.~\ref{subsubsec:grad}, we apply it to the data, to recover the time evolution of the central metallicity and the MW metallicity gradient. 

\begin{figure*}
    \centering
    \includegraphics[width=0.8\textwidth]{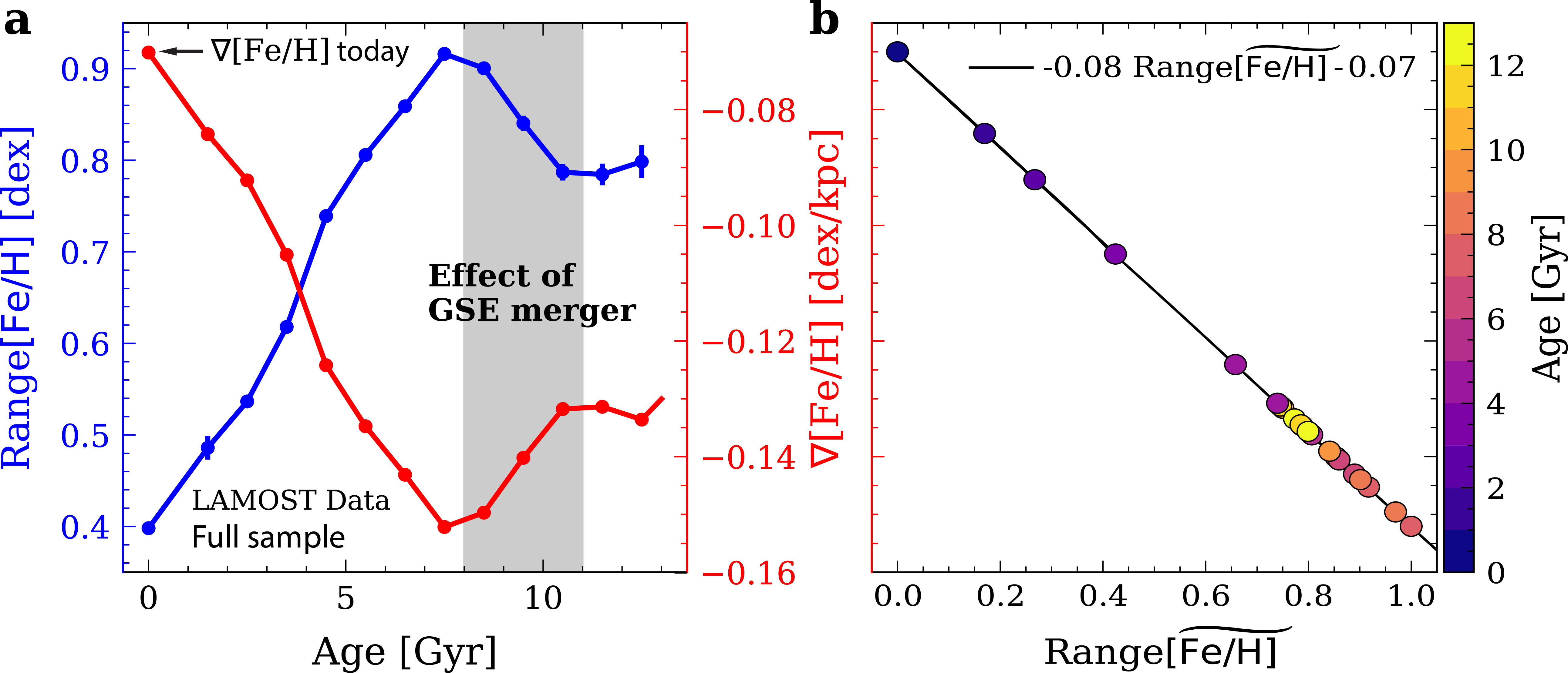}
    \caption{Recovering the birth metallicity gradient directly from the data. a: The metallicity range of our sample (blue curve) as a function of stellar age in 1-Gyr-wide bins. The time evolution of the birth $\mathrm{\nabla [Fe/H]}$ (red curve) is expected to anti-correlate with the current day Range[Fe/H]. The gray vertical strip indicates the steepening of the gradient, which coincides with the time of the MW's last massive merger, GSE. b: We impose a linear relation between $\mathrm{Range\widetilde{[Fe/H]}}$ (normalized Range[Fe/H]) and $\mathrm{\nabla [Fe/H]}$, in agreement with cosmological simulations of disk formation (see Fig.~\ref{fig:1}). The y-intercept in the line equation shown in this panel is fixed by the present day gradient $\mathrm{\nabla [Fe/H](\tau=0)}=-0.07$~dex/kpc \citep{Braganca19}. The line slope of $-0.08$ dex/kpc defines the overall gradient steepness, and is set by requiring that the youngest stars in the solar neighborhood are born locally. See Sec.~\ref{subsection:method} for more details. 
    }
    \label{fig:fehevo}
\end{figure*}

\subsection{Insights from simulations: [Fe/H] range today as a tracer of the birth gradient evolution}\label{subsubsec:val}

Even though the birth metallicity gradient as a function of cosmic time in the MW is now lost, we can gain insights from cosmological simulations of MW-like galaxies on how it may relate to current day observables. We propose here that the metallicity scatter, or range, as a function of age correlates well with the time evolution of the birth metallicity gradient, $\mathrm{\nabla [Fe/H]}(\tau)$. 

Such an inverse correlation can be expected. An extreme example would be a completely flat radial birth metallicity gradient, in which all stars formed would have the same [Fe/H], thus causing the range in [Fe/H] to be 0. On the other hand, stars forming along a steeper metallicity gradient would create a larger range of [Fe/H], since the stars throughout the disk are forming with many different values of [Fe/H].

We tested this idea using four simulations of disk galaxies in the cosmological context from two simulation suites (see Sec.~\ref{sec:sims}), including barred and non-bared galaxies, and obtained very similar results. Two of these simulations, both similar to the MW, are presented in Fig.~\ref{fig:1}. In the top row we show galaxy g2.79e12 from the NIHAO-UHD project \citep{Buck2020}, which has been studied extensively and known to have a massive merger at the beginning of disk formation, similar to GSE for the MW. The bottom row of Fig.~\ref{fig:1} presents the MW analog from the 17\_11 HESTIA simulation \citep{Libeskind2020, Khoperskov2022}.

For each simulation shown in Fig.~\ref{fig:1}, we use age bins of width 1~Gyr, (as in the data) and only select stellar particles that have present day radii between 7 and 10~kpc, in order to mimic the radial range of the data where most stars are. We used 1 Gyr as the age bin size for the data and simulation not only because we only selected stars with age uncertainty $<$ 1 Gyr, but also because when a merger comes in, it disrupts the linearity of the metallicity-\rbir\ relation for $\sim$ 200-400 Myr based on simulations \citep{Lu2022c}. We account for the measurement uncertainties by perturbing the metallicity and age for each star particle 100 times, based on the typical uncertainty of the data ($\sigma_{\mathrm{[Fe/H]}} = 0.03$ dex; $\sigma_{\tau} = 0.50$ Gyr; see below). To account for outliers, we approximate the [Fe/H] range as the 95\%-5\%-tile of the metallicity distribution in a given age bin, namely Range[Fe/H] = 95\%[Fe/H]-5\%[Fe/H], for stars in each 1 Gyr age bin.

We estimate the error in this measurement (given the imposed age and [Fe/H] error), by making the same calculation but using the perturbed metallicity and age within their uncertainties 100 times. Range[Fe/H] and its uncertainty are reported as the average and standard deviation of these 100 runs, respectively. 

The true birth metallicity gradient $\mathrm{\nabla [Fe/H]}$ is calculated using an $L^2$ minimization for all the stars born in-situ (by selecting stars with [Fe/H] $>$ -1 dex and excluding those with eccentricity $>$ 0.5 and galactic height $>$ 1 kpc) within a lookback time equal to that age bin. We see in Fig.~\ref{fig:1}a,c that for both simulations the $\mathrm{Range[Fe/H]}$ variation with age (blue curve) and the $\mathrm{\nabla [Fe/H]}$ variation with lookback time (red solid curve) anti-correlate very well. In panels b and d we find a strong linear relation (Pearson's Correlation Coefficient, PCC $= -0.97$ and $-0.98$ for the NIHAO-UHD and the HESTIA simulations, respectively) between the two functions. 

The tilde sign in $\mathrm{Range\widetilde{[Fe/H]}}$ indicates that the Range[Fe/H] has been normalized to lie between 0 and 1, which now represents the gradient {\it shape function}, and is unitless. The linear relation seen in Fig.~\ref{fig:1}b,d can be written as
 \begin{equation}\label{eq2}
    \mathrm{\nabla [Fe/H]}(\tau) = a\ \mathrm{Range\widetilde{[Fe/H]}(age)}+b,
\end{equation}
where the scale factor $a$ controls the overall strength of the gradient and $b$ gives the present day value; for example, $a=-0.09$ dex/kpc and $b\equiv\mathrm{\nabla [Fe/H]}(\tau=0)=-0.02$~dex/kpc for NIHAO-UHD, as seen in panel b. The steepest $\mathrm{\nabla [Fe/H]}$ value is $a+b=-0.11$ dex/kpc, agreeing with the minimum of the red curve in panel $a$. 

The above equation illustrates how a variable dependent on age (Range[Fe/H]) is transformed into a variable dependent on lookback time ($\mathrm{\nabla [Fe/H]}$). It is worth stressing again that the metallicity gradient at lookback time, $\tau$, is not the same as the metallicity gradient measured from that mono-age population (hence why we distinguish $\tau$ and age in Eq.~\ref{eq2}), as radial migration flattens the birth gradient at lookback time significantly (see Sec.~\ref{sec:intro}). 

To find out what uncertainty results from imposing an exact linear relation between the range and the radial birth gradient, we calculate the {\it inferred} birth gradient by shifting the points in Fig.~\ref{fig:1}b,d verticaly to lie exactly along the red line. The result is plotted as the dashed red curve in panels a and c, showing that the uncertainty is, on average, 9\% (NIHAO-UHD) and 7\% (HESTIA) of the true gradient (solid red curve). We also checked that changing the stellar particle radial range from 7-10 kpc to 8-9 kpc or to 6-11 kpc affects our results minimally.

\subsection{Recovering the MW birth metallicity gradient evolution with time}
\label{subsubsec:grad}

With the above insights from the simulations, we can now derive the Galactic birth metallicity gradient evolution with cosmic time, $\mathrm{[Fe/H]}(R_b,\tau)$, in a relatively straightforward manner directly from the data.

We start by calculating the metallicity range as we did for the simulations. This is shown as the blue curve in Fig.~\ref{fig:fehevo}a, while the red curve is the [Fe/H] birth gradient estimated from its inverse relation to the range, as shown in Fig.~\ref{fig:fehevo}b and modeled as follows.

If [Fe/H] is always linear in $R$, we can write for any lookback time $\tau$, 
\begin{equation}\label{eq3}
    \begin{split}
    \mathrm{[Fe/H]}(R_b,\tau) = \mathrm{\nabla [Fe/H](\tau)} R_b + \mathrm{[Fe/H]}(0,\tau)=\\
    = (a\ \mathrm{Range\widetilde{[Fe/H]}(age)}+b) R_b + \mathrm{[Fe/H]}(0,\tau),
    \end{split}
\end{equation}
where in the second line we have substituted Eq.\ref{eq2} for the time evolution of the gradient.
Here the constant $b\equiv \mathrm{\nabla [Fe/H]}(\tau=0)\approx -0.07$~dex is the present day radial metallicity gradient in the MW \citep{Braganca19}, and the scale factor $a$ is a constant defining the overall gradient steepness, and is to be determined. 
Lastly, $\mathrm{[Fe/H]}(0,\tau)$ is the time evolution of the metallicity at the Galactic center, which is also to be determined.

\begin{figure}
    \centering
    \includegraphics[width=0.4\textwidth]{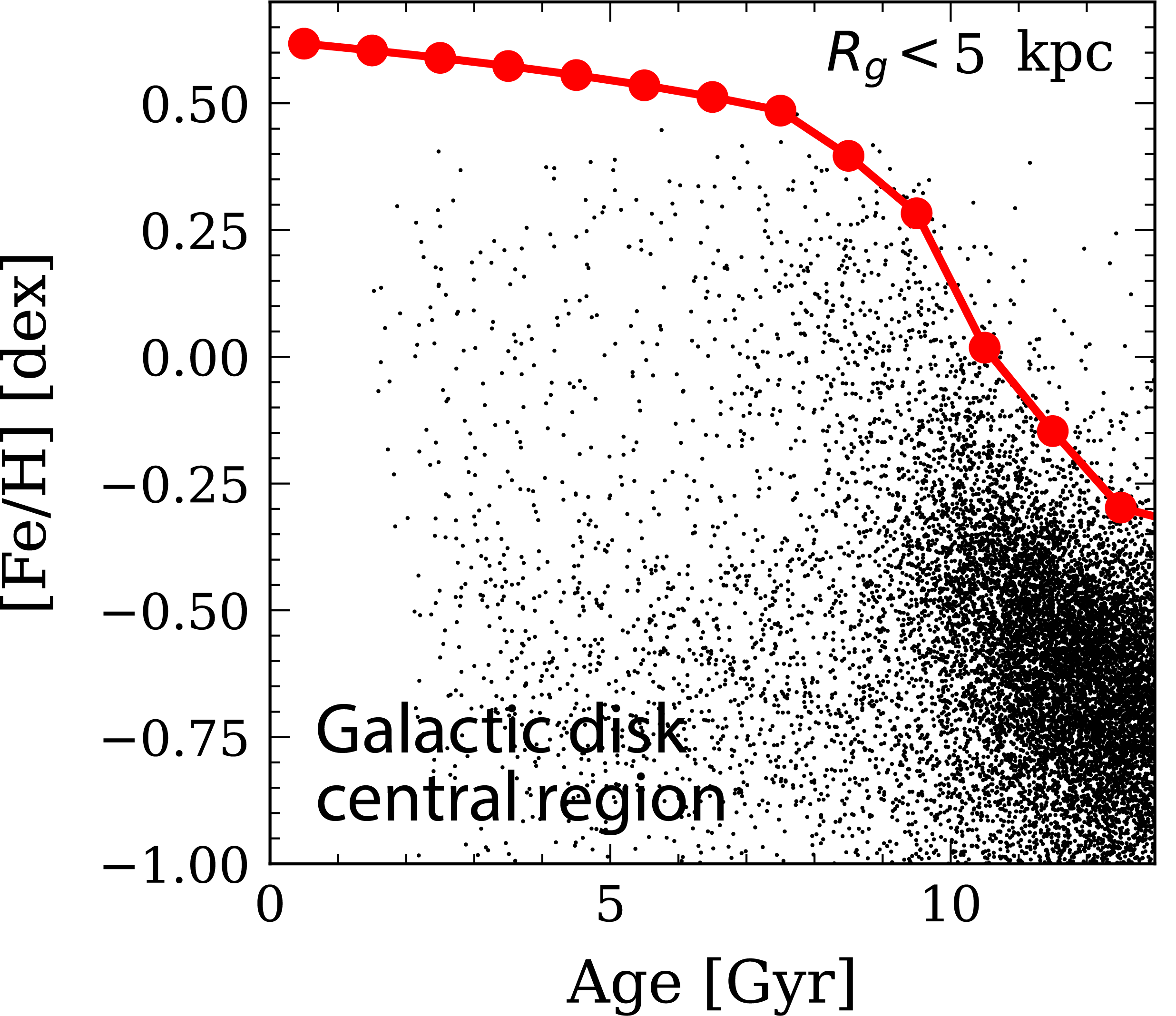}
    \caption{Estimating the central metallicity time evolution.
    Scatter shows the AMR of stars currently found at $R_g<5$~kpc. The metallicity enrichment of the central disk region, $\mathrm{[Fe/H]}(0,\tau)$ (last term in Eq.~\ref{eq3}) for ages $>7$~Gyr, is estimated from the upper boundary of the AMR (red curve). For age $<7$~Gyr we increase monotonically $\mathrm{[Fe/H]}(0,\tau)$ over time, using a log function, at a rate that matches the $\sim$0.1 dex metallicity of the youngest stars at the solar neighborhood \citep{nieva12}.}
    \label{fig:2}
\end{figure}

\subsubsection{Inferring the central metallicity time evolution in the MW}

To estimate $\mathrm{[Fe/H]}(0,\tau)$ - the last term in Eq.~\ref{eq3}, in Fig.~\ref{fig:2} we plot the AMR for stars currently found in the inner MW disk, with guiding radii $R_g<5$~kpc. 
The exponential density drop with radius and inside-out formation of disk galaxies suggest the central region should have the highest star formation rate and therefore, has the highest metallicity at all lookbak times.
As a result, the upper envelope in this selection should represent the central metallicity evolution with time, assuming the metallicity gradient was always negative \citep{Hemler2021}.

The red curve in the same plot traces the AMR upper boundary for age $>7$~Gyr, estimated as the 95\%-tiles in metallicity for stars in 1-Gyr wide bins. Due to the lack of younger stars in this central region, for age $<7$~Gyr we fit a log function so that $\mathrm{[Fe/H]}(0,\tau)$ monotonically increases with time at a rate that matches the $\sim$0.1 dex metallicity of the youngest stars \citep{nieva12} in the solar neighborhood, $\mathrm{[Fe/H]}(8.2,0)=0.1$~dex. 
In the future, the true central metallicity evolution can be estimated (especially for stars $<$ 7 Gyr) as more ages and metallicity measurements become available from SDSS-V \citep{Kollmeier2017}.

\subsubsection{$\mathrm{\nabla [Fe/H]}$ strength determination}
\begin{figure*}
    \centering
    \includegraphics[width=1\textwidth]{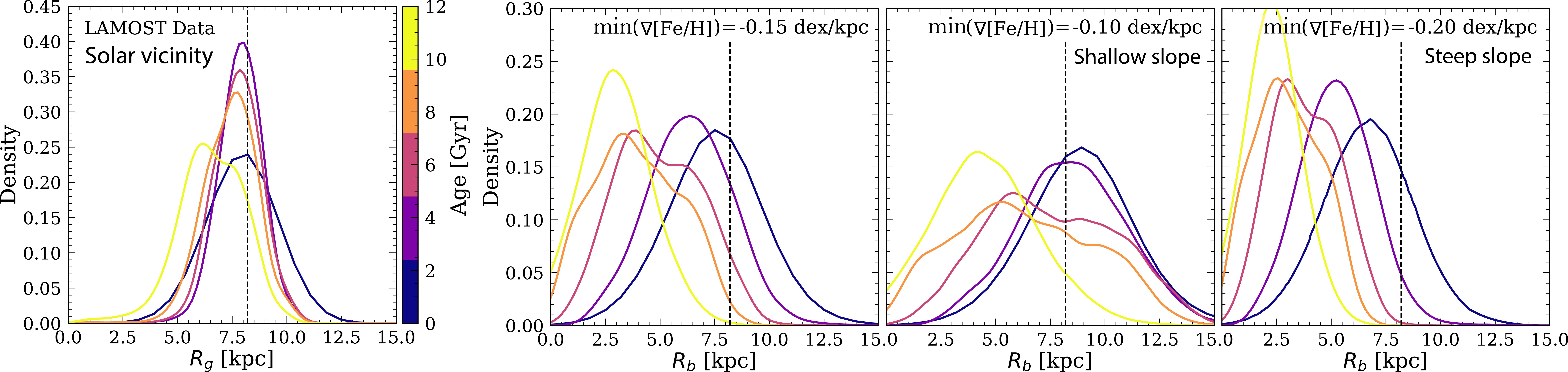}
    \caption{The effect of gradient strength on derived \rbir. First panel: Distribution of guiding radius $R_g$ for stars in the solar vicinity (7.7 kpc $<$ R $<$ 8.7 kpc) in bins of age. The area under each curve is normalized to one. Second panel: \rbir\ distributions for mono-age populations, as derived from Eq.~\ref{eq3} using a scale factor $a=-0.08$ dex/kpc, which results in $\mathrm{min(\nabla [Fe/H]})\sim-0.15$~dex/kpc. Making this shallower ($-0.1$~dex/kpc; Third panel) results in very wide \rbir\ distributions, while a steeper gradient ($-0.2$~dex/kpc; Fourth panel panel) suggests the distribution peak of the youngest population has shifted outwards by about 1.5~kpc. Both of these are not expected \citep{Minchev2018}.
    }
    \label{fig:4}
\end{figure*}

We constrain the overall gradient steepness -- the scale factor $a$ in Eq.~\ref{eq3}, via an iterative process by examining its effect on our birth radius estimates. 
In Fig.~\ref{fig:4} we show the \rbir\ distributions of mono-age groups for the solar neighborhood sample, using different values for $a$ in Eq.~\ref{eq3}. In the second panel we can see the results for $a=-0.08$ dex/kpc, corresponding to $\mathrm{min(\nabla [Fe/H]})\sim-0.15$~dex/kpc, for which the youngest mono-age population peaks very close to its current position (as seen from the guiding radius distributions in the leftmost panel). The slightly inward shift is expected \citep{Minchev2018} given the very few stars with age $<1.5$~Gyr in our LAMOST sample. In contrast, changing the scale factor so that the steepest gradient is $\sim-0.1$~dex/kpc (third panel) results in the youngest stars having an \rbir\ peak outside the solar radius; this is unphysical since due to the exponential density drop of the disk and its inside-out formation, stars in all mono-age populations will preferably shift outwards as there are more stars born in the inner Galaxy compared to the outskirt \citep{Roskar2008, Minchev2013,Ma2017}.
This means, the birth radii distribution for stars of a mono-age population will peak at a smaller radius than the current day radii distribution.
Finally, setting $\mathrm{min(\nabla [Fe/H]})\sim-0.2$~dex/kpc shifts the youngest \rbir\ distribution more than a kpc inwards (rightmost panel), which cannot be justified, keeping in mind that stars require time to migrate (e.g., \citealt{Frankel2020}).

The blue curve in Fig.~\ref{fig:fehevo}a shows the present-day metallicity range, Range[Fe/H], measured from our sample, as a function of age.
The red curve shows the derived birth radial metallicity gradient, $\mathrm{\nabla [Fe/H]}(\tau)$, as a function of lookback time using the method described in this Sec.. Table~\ref{tab:1} lists the $\mathrm{\nabla [Fe/H]}(\tau)$ and $\mathrm{[Fe/H]}(0,\tau)$ values in 1-Gyr bins.

\begin{table}
    \centering
    \begin{tabular}{c|c|c}
         $\tau$ [Gyr] &  $\mathrm{\nabla [Fe/H]}(\tau)$ [dex/kpc] & [Fe/H](0, $\tau$) [dex]\\
         \hline
         0 & -0.070 & 0.624 \\
         0.5 & -0.075 & 0.618 \\
         1.5 & -0.084 & 0.604 \\
         2.5 & -0.092 & 0.588  \\
         3.5 & -0.104 & 0.570 \\
         4.5 & -0.124 & 0.549 \\
         5.5 & -0.135 & 0.524 \\
         6.5 & -0.143 & 0.493 \\
         7.5 & -0.152 & 0.451 \\
         8.5 & -0.150 & 0.396 \\
         9.5 & -0.140 & 0.283 \\
         10.5 & -0.132 & 0.018 \\
         11.5 & -0.131 & -0.147 \\
         12.5 & -0.133 & -0.297 \\
         13.0 & -0.130 & -0.315 \\
    \end{tabular}
    \caption{Time evolution of the birth radial metallicity gradient and the central disk metallicity.}
    \label{tab:1}
\end{table}

\subsection{Estimating \rbir\ for the LAMOST sample}
\label{sec:rb}

To estimate the birth radius we simply express \rbir\ in terms of the measured stellar age and [Fe/H] in our sample, using Eq.~\ref{eq3}:
\begin{equation}\label{eq4}
    \mathrm{R_b(age, [Fe/H])} = \frac{\mathrm{[Fe/H]} - \mathrm{[Fe/H]}(0,\tau)}{-0.08\ \mathrm{Range\widetilde{[Fe/H]}(age)}-0.07}.
\end{equation}
In the above, the functional form of $\mathrm{[Fe/H]}(0, \tau)$ can be found in Table~\ref{tab:1}, we have set for the scale factor $a=-0.08$ dex/kpc as derived above, the constant $b=-0.07$ dex/kpc is given by the present day gradient derived using OB stars from \cite{Braganca19}, and the $\mathrm{Range\widetilde{[Fe/H]}(age)}$ was measured from the data. 
To check the effect of the current day MW metallicity gradient measurement, instead of -0.07 dex/kpc (the constant b from Eq.\ref{eq2}), we used -0.04 dex/kpc and found an overall shift in absolute \rbir, but no significant change in its relative value. We found, therefore, no major changes to the main results.

\begin{figure}
    \centering
    \includegraphics[width=0.4\textwidth]{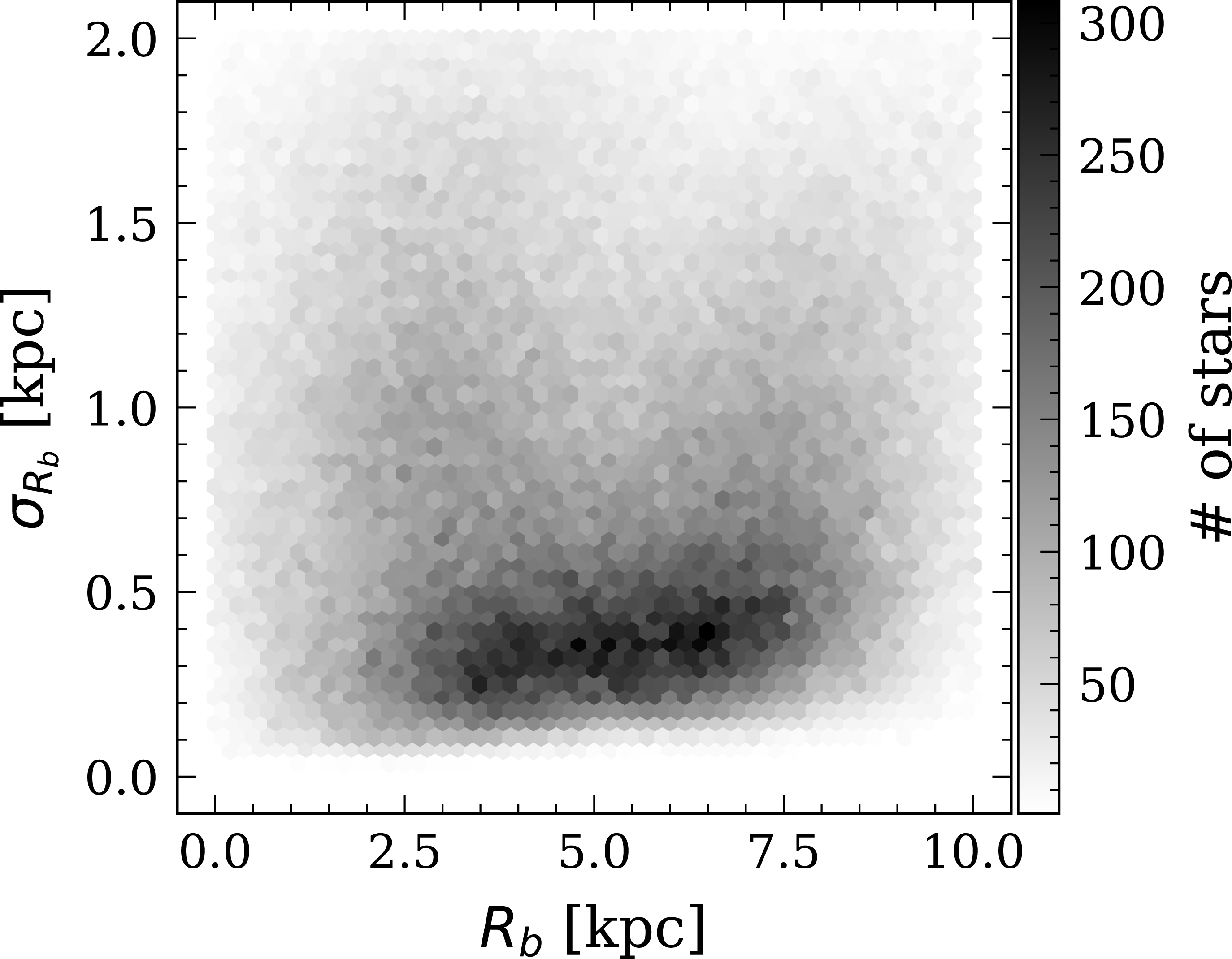}
    \caption{\rbir\ versus the \rbir\ error, $\sigma_{R_b}$ for 217,672 subgiant stars from our data set \citep{Xiang2022}, for age $<13$ Gyr and [Fe/H] $>-1$ dex. The median error is 14\% or 0.72 kpc and the uncertainty is estimated using bootstrapping.}
    \label{fig:5}
\end{figure}

We use Eq.~\ref{eq4} to estimate \rbir\ for 217,672 subgiant stars \cite{Xiang2022} with age $<13$ Gyr and [Fe/H] $>-1$ dex. The uncertainties $\sigma_{R_b}$ are estimated by perturbing the age and metallicity within their errors and recalculating \rbir. We performed this bootstrapping 100 times and used the standard deviation as the \rbir\ error. The stellar distribution as a function of \rbir\ and $\sigma_{R_b}$ is shown in Figure~\ref{fig:5}. 
The majority of stars are seen to have $\sigma_{R_b}<0.5$~kpc, or a median uncertainty of 13\%. 

To test how well this method works, in Sec.~\ref{sec:rb_test} we applied it to the g2.79e12 simulation by using just the last snapshot, finding that we could recover \rbir\ within 20\% (Figure~\ref{fig:A5}), as well as the structure of the AMR and the \alphafe-\feh\ plane in terms of \rbir\ (Figure~\ref{fig:A6}).

\section{Results}\label{sec:results}
\subsection{Time evolution of the MW metallicity gradient}\label{subsec:feh}
Unlike in conventional forward chemo-dynamical modeling, our empirically derived time evolution of the MW metallicity gradient naturally takes into account the net effect of any and all physical processes affecting the disk's chemical evolution, e.g., the effects of past merger events, such as GSE \citep[e.g.][]{Belokurov2018, Helmi2018} and the Sagittarius dwarf galaxy \citep{Ibata1994}, as well as gas flows \citep{Lacey1985} and Galactic fountains \citep{fraternali17, Marasco2022}. 
Fig.~\ref{fig:fehevo}a shows that the MW exhibits a significant radial metallicity gradient (red curve) early on, as often seen in simulations after a MW-like disk has just started to form \citep{Hemler2021, Lu2022}, which for the Galaxy can be as early as 12-13 Gyr ago \citep{Conroy2022,Rix2022}. Interestingly, we find a steepening of the gradient from $-0.13$ to $-0.15$ dex/kpc, over a period of about 3 Gyr (from 11 to 8 Gyr ago; vertical gray strip). This transformation coincides with the conclusion of a significant merger event in MW's history -- the GSE merger that happened 8-10~Gyr ago \citep{Belokurov2018,Helmi2018} or even one Gyr earlier \citep{Xiang2022}. 
At lookback time less than 8 Gyr, we recover a monotonically decreasing gradient up to the present-day value.

Recently, \cite{Buck2023} investigated the steepening of the metallicity gradient in the g2.79e12 simulation from NIHAO-UHD (seen also in our Fig.~\ref{fig:1}) and concluded that it could be caused by a rapid increase in the cold gas surface density of the galaxy outskirts from a gas-rich merger, breaking the self-similar enrichment of the inter-stellar-medium.

\begin{figure*}
    \centering
    \includegraphics[width=0.8\textwidth]{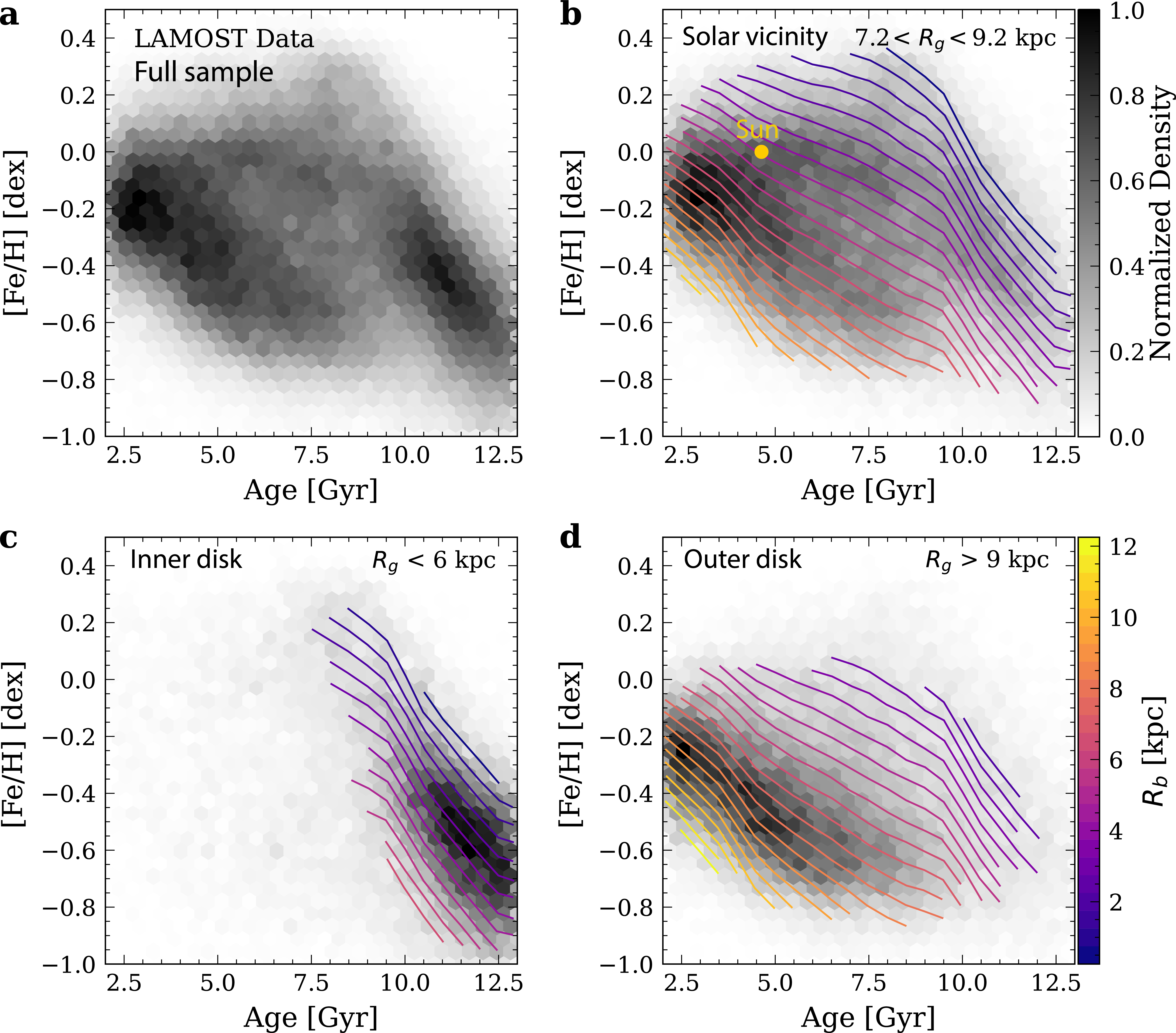}
    \caption{Assembly of the MW age-metallicity relation (AMR).
    a: Stellar density distribution of the full data set, shown as a 2D histogram in bins of $\Delta$\feh=0.05 dex and $\Delta$age=0.37 Gyr. Darker colors correspond to higher density. The sample is further split into stars with guiding radius, $R_g$, currently located in the solar vicinity (b), the inner disk (c), and the outer disk (d). The curves overlaid on top of panels b,\ c,\ d show the AMR of different birth radii using a bin of $\Delta$\rbir=0.5~kpc.
    It now becomes clear that the ridge of old stars originated in the inner 5~kpc, while the younger ridge is composed of stars mostly born outside 6~kpc. Because of radial migration and the asymmetric drift, both of these are present in the solar vicinity.
    Given the Sun's age and metallicity (yellow dot in panel b), we estimate that it was born at $R=4.5\pm 0.4$~kpc. See Sec.~\ref{sec:rb} for more details.}
    \label{fig:amr}
\end{figure*}

\subsection{Formation of the MW age-metallicity relation (AMR)}\label{subsec:amr}
With the addition of \rbir\ to our sample, we can now better understand the structure of the AMR, which tells us how the Galactic disk enriched with iron over time.
Fig.~\ref{fig:amr}a shows the density distribution of the total sample, with a darker color corresponding to a higher density. Two well-defined ridges with negative slopes are apparent for younger and older stars as seen in \citep[e.g.][]{Xiang2022, Sahlholdt2022}. In Fig.~\ref{fig:amr}b,c,d we show the stars with current guiding radius $R_g$ near the Sun, in the inner disk, and the outer disk, respectively. It is remarkable that the bimodal structure seen in the total sample splits cleanly, in that the older ridge belongs to the inner disk (panel c) and the younger one to the outer disk (panel d), while the Sun falls in the transitional region (panel b). This already indicates that the prominent overdensities found in the total sample are in fact due to a smooth transition from the inner to the outer disk, rather than two independent disk components.

To find out exactly where the AMR ridges originated, in Fig.~\ref{fig:amr}b,c,d we overlay the running mean of stars born at different \rbir, as color-coded.
The \rbir\ curves match remarkably well the slopes of the two over-densities – steeper for the old ridge and shallower for the younger. Indeed, the two structures are associated with very different disk radii. 
A strong star-formation rate early on can naturally explain the older ridge over-density \citep[][]{Xiang2022}, reaching the maximum [Fe/H] values in our sample at the end of the $\mathrm{\nabla [Fe/H]}$ steepening period ($\sim8$~Gyr, left edge of gray strip). Thanks to our \rbir\ estimates we can see that these stars have formed in the inner 4-5~kpc, indicating the extent of the disk 8-10 Gyr ago. On the other hand, the younger ridge starts forming at the end of the $\mathrm{\nabla [Fe/H]}$ steepening period at age $<$ $\sim$8 Gyr and at \rbir$>$~5-6~kpc.
This ridge is possibly formed from gas brought in by the GSE merger \citep[e.g.][]{Buck2023}. The slower metallicity increase with time in this feature indicates a dynamically quiescent Galactic disk formation epoch \citep[also pointed out in][]{Xiang2022}. 
The over-density of the young ridge can be explained by the densely spaced \rbir\ curves along this ridge, assuming a relatively constant star formation with radius following the last massive merger.

Another interesting feature in Fig.~\ref{fig:amr} is the decrease in metallicity of the AMR upper boundary, which can be misinterpreted as a chemical dilution in the last 8 Gyr. A more natural explanation, in light of the \rbir\ estimates, is that the decrease results from the superposition of different \rbir\ tracks. 
Due to selection effect, the youngest, most metal-rich stars born in the inner Galaxy haven't had time to migrate to the solar vicinity, and thus, are not observed.
This is an example of Simpson’s paradox -- a commonly encountered statistical phenomenon in the field of Galactic Archaeology \citep{Minchev2019}.

 \begin{figure*}
    \centering
    \includegraphics[width=.9\textwidth]{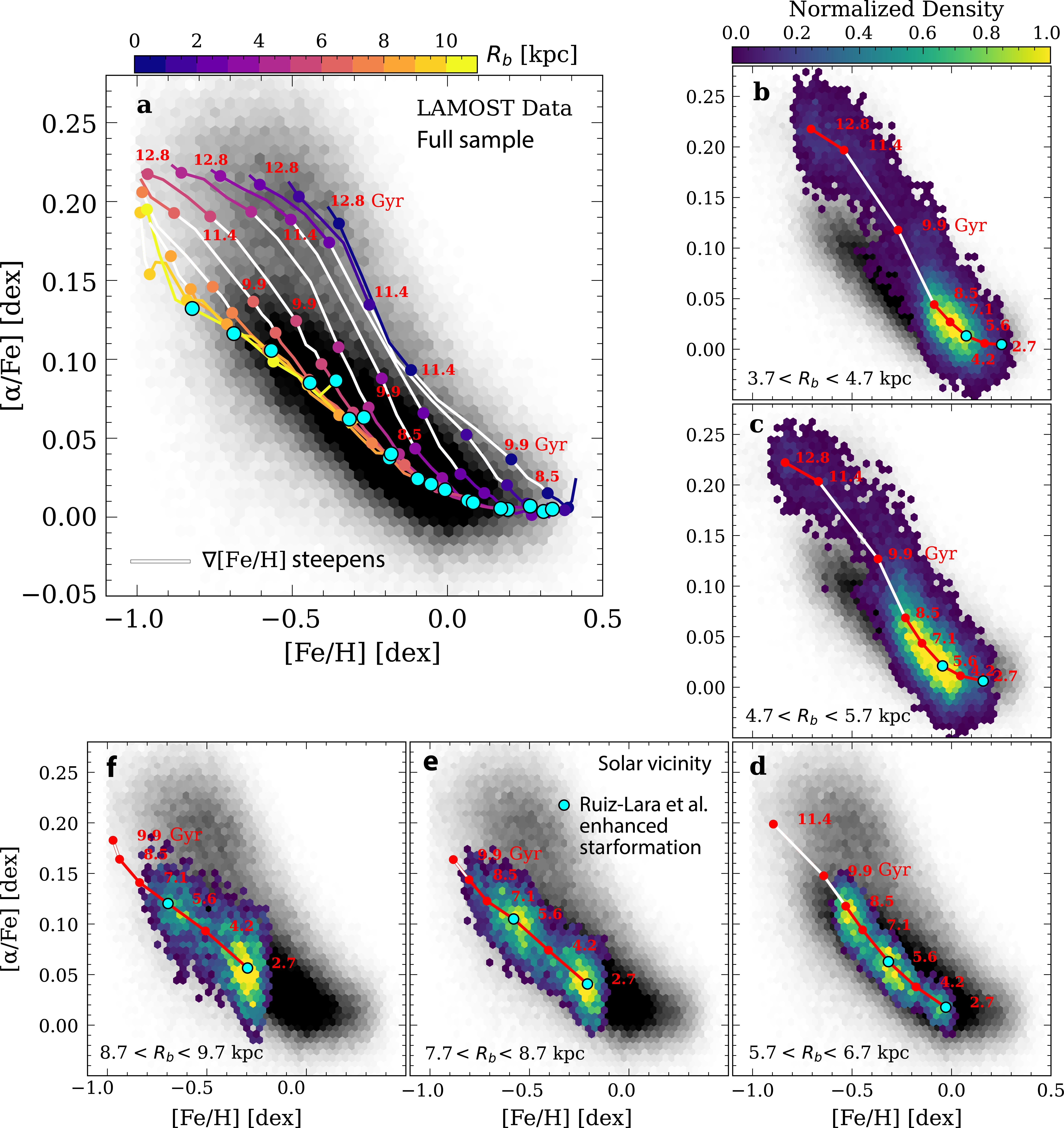}
    \caption{Formation of the \alphafe-\feh\ bimodality.
    The background in each panel shows the stellar number density for the full sample. a, The color-coded curves trace the time evolution of stars born in mono-\rbir\ bins, as indicated in the colorbar on top. Each curve is produced by binning a mono-\rbir\ population ($\Delta$\rbir=1~kpc) by age and estimating the mean \alphafe\ and \feh\ values in each age bin, $\Delta$age=1.4~Gyr. Note that when working with \rbir, age becomes lookback time, $\tau$.
    The white segment in each \rbir\ track corresponds to the time during which $\mathrm{\nabla [Fe/H]}$ is found to steepen (from 11 to 8~Gyr ago; gray strip in Figs.~\ref{fig:fehevo} and \ref{fig:amr}), coinciding in time with the GSE merger impact. This period of time outlines well the gap between the two sequences, thus suggesting the GSE is responsible for its formation. The dots along each curve are separated uniformly in time (every 1.4 Gyr) with the largest distance traveled across the white segment. This indicates that the gap results from the quick drop in \alphafe\ on a Gyr timescale.
     b-f, Colored contours show the \alphafe-\feh\ relation for a 1-kpc-wide \rbir\ bin, for five different radii, as indicated in the bottom of each panel. The red curves track the time evolution as in panel a. For \rbir$\gtrsim6$~kpc (panels d,e,f) we find two well-defined clumps of mean age (cyan dots) very similar to a recent estimate of enhanced star formation episodes \citep{RuizLara2020}. Although the clumps shift position in this plane for different \rbir, they are still found at the same mean age. This figure shows the power of having \rbir\ at our disposal, as the structure we see is washed out in the total population.
    }
    \label{fig:afefeh}
\end{figure*}

\subsection{The \alphafe\ bimodality}\label{subsec:alphafe}
One of the most intriguing findings in spectroscopic data is the dichotomy of the stellar distribution in the [$\alpha$/Fe]-[Fe/H] plane \citep{Hayden2015, Queiroz2020}. This is shown for our data in Fig.~\ref{fig:afefeh}a as the gray-scale 2D histogram of the stellar number density. The high-\alphafe\ sequence (clump above [$\alpha$/Fe]$\sim0.15$~dex but extending to lower values at higher [Fe/H]), is mostly old, while the low-\alphafe\ sequence is known to have a large spread in age.

Many models of how the bi-modality of the \alphafe-[Fe/H] was formed make sets of specific predictions.
One example is the two-infall model \citep[e.g.][]{Chiappini1997, Spitoni2021, spitoni2022}.
This model consists of two main infall episodes where the thick disc forms fast by a fast gas accretion event, and the thin disc forms by a second accretion episode on a longer time-scale.
The two-infall model is able to predict the bi-modality of \alphafe-[Fe/H], age-metallicity relation, stellar metallicity distribution, and many other features presented in the observations.
However, the two-infall model failed to explain the most metal enriched \ha~stars in the \alphafe-[Fe/H] plane \citep[e.g.][]{Matteucci2021A,Grisoni2017}.
These stars can be explained only as stars migrated from the inner Galactic regions.
Simulations from both \cite{Agertz2020} and \cite{Buck2020} resemble the two-infall model and suggest that the $\alpha$-bimodality is a generic consequence of gas-rich mergers diluting the metallicity of the gas in the high-$\alpha$\ disk, forming the \lowalpha\ disk.

One other model is the parallel model, where the high- and \lowalpha\ disks are treated as two distinct evolutionary phases that evolve separately with separate star formation rates (SFR).
The clumpy star formation scenario closely resembles such a scenario \cite[e.g.][]{Clarke2019, Debattista2019}.
Star forming clumps dominate in many galaxies and MW progenitors for readshift $>$ 2 \citep[e.g.][]{Elmegreen2005,Elmegreen2007,Cowie1995,van1996}.
In the clumpy star formation model, the \ha~disk stars are formed from clumps with high SFR, and the \lowalpha~disk stars are created via low SFR gas. Other scenarios are also proposed such as radial migration \citep{Sharma2021}.
For a detailed review, see Sec. 5 from \cite{Matteucci2021A}.

By adding in the dimension of \rbir\ to the age and abundance measurements of our data, we are now able to fully understand this relation. 
The color-coded curves in Fig.~\ref{fig:afefeh}a trace the time evolution of stars born in mono-\rbir\ bins in the \alphafe-\feh\ plane. 

We find that the high-\alphafe\ sequence has formed mostly at \rbir$<5$~kpc.
It then extends to higher [Fe/H] and lower \alphafe\ along the innermost \rbir\ tracks (blue curves), and the transition takes place before and partly during the $\mathrm{\nabla [Fe/H]}$ steepening period (shown by the white segments of the \rbir\ tracks). 
These attributes indicate strongly that the high-\alphafe\ sequence is composed of the same stars that form the older [Fe/H] ridge in the AMR \citep[Fig.~\ref{fig:amr}; also shown in][]{Chiappini1997, Matteucci2021, Sahlholdt2022}. 
Given the fast chemical enrichment over a very short timescale, both of these features must have formed in a gas-rich medium with very efficient star formation, in agreement with previous work \citep{Chiappini1997, Sharma2021}. Here, however, we are able to track in detail their evolution with both birth radius and lookback time directly from the data. 
On the other hand, the low-\alphafe\ sequence can in turn be matched to the younger ridge in the AMR seen in Fig.~\ref{fig:amr}, given the common \rbir\ extent, time span, and metallicity range of these two structures. The wider gap between the two ridges in the AMR compared to the \alphafe-\feh\ plane, and the strong overlap of birth radii in the low-\alphafe\ sequence, are both due to the convoluted relation between age and \alphafe, as they correlate well only for mono-\rbir\ populations (see, e.g., \citealt{Minchev17, Ratcliffe23b}).

A remarkable feature in Fig.~\ref{fig:afefeh}a is the fast transition between the high- to low-\alphafe\ disks seen for the innermost $\sim6$ kpc (blue/purple curves). The gap between the two sequences originates from the quick time transition between them -- the dots, regardless of color, are separated by 1.4~Gyr and the spacing across the gap is significantly larger than the rest. This presents evidence for inside out formation, already taking place within this very concentrated disk very early on (about 5 kpc in radius and 11 Gyr ago) -- star formation in the innermost few kpc starts earlier, thus reaching sooner the Type Ia Supernova dominated evolutionary phase that causes the fast drop in \alphafe. Moreover, the white segment in each \rbir\ track indicates the time period during which $\mathrm{\nabla [Fe/H]}$ steepens (gray strips in Figs.~\ref{fig:fehevo} and \ref{fig:amr}). This means, the transition between the high- and low-\alphafe\ disk could be a combination of Type Ia Supernova delay and the GSE merger. 

The \rbir\ tracks of the inner 5-6 kpc transition smoothly from the high- to the \lowalpha\ sequence, but only stars born in the outer radii ($> \sim$ 8 kpc) contribute to the very metal-poor end of the \lowalpha\ disk (at [Fe/H] $\lesssim-0.4$ dex). 
These stars are born during or after the GSE merger, indicating that the most metal-poor \lowalpha\ disk stars are most likely born directly from the gas brought in from the GSE merger.

To test how uncertainty can affect the robustness of our results from Figures~\ref{fig:amr} and \ref{fig:afefeh}, we convolved synthetic error into \rbir, drawing from a Gaussian distribution with different widths. We found that the features remained significant until the \rbir\ uncertainty reached $\sim30\%$.

\subsection{Discovery of coeval clumps in the \alphafe-\feh\ plane}\label{subsec:clumps}

In Fig.~\ref{fig:afefeh}b-f we show the 2D stellar density in the \alphafe-\feh\ plane for five mono-\rbir\ populations. Smooth variation is found in the inner $\sim6$~kpc, however two well-defined clumps appear in each of the three outer disk bins, including the solar vicinity. These over-densities have the same mean ages in all three \rbir\ bins (marked by the cyan dots), which are very similar to the times of two known star-formation busts associated with the Sagittarius dwarf galaxy \citep{RuizLara2020, Laporte2018}. For the first time, we are here able to see structure in the low-\alphafe\ sequence, which is otherwise washed out by radial migration.

To check the robustness of the above described features, we considered the effect of the uncertainty that can result in the determination of $\mathrm{\nabla [Fe/H]}$. To this end, we perturbed the inferred metallicity gradient (see Fig.~\ref{fig:fehevo}) by 10\% with Monte Carlo sampling 100 times and recalculating \rbir. We then estimated the \rbir\ uncertainty as the standard deviation of these 100 outcomes. 
It was found that Figures~\ref{fig:amr} and ~\ref{fig:afefeh} were mostly unchanged, ensuring that our main conclusions are robust.

\subsection{The Solar birth radius}

The Sun has been suggested to originate from inside the solar circle since the work by \cite{wielen96}, given its higher metallicity compared to local stars of similar age. Using [Fe/H] $=0\pm0.05$~dex \citep{Asplund2009} and age $=4.56\pm0.11$~Gyr \citep{Bonanno2002} in Eq.~\ref{eq4}, we estimate $\mathrm{R_{\odot, b}}=4.5\pm0.4$~kpc. This value is one of the lowest estimates in the literature \citep{wielen96, Minchev2013, Ratcliffe23b, baba23}.

\section{Limitations}\label{subsec:limiation}

Since the detailed formation history of the MW is unknown, simulations and high-redshift MW-like galaxies are the only few pathways to understand that. While disk formation simulations in the cosmological context provide important insights into the MW evolution, unknown subgrid physics, limitations in resolution, different feedback mechanisms, and chemical enrichment prescriptions (e.g., yields) may affect our results. 

Notwithstanding these limitations, the fact that two different suites of simulations both produce similar results, suggests that the metallicity enrichment is less subjective to the subgrid prescriptions, providing supporting evidence for the method developed in this work.

We also acknowledge that the metallicity gradient at birth may not have been linear throughout the entire MW evolution \citep[e.g.][]{Sanchez2014, Braganca19, Luck2018}. From the simulations we tested, this assumption is true after the stellar disk has started to form \citep[e.g.][]{Lu2022}. 

Finally, we would like to mention that this method requires precise and accurate ages for stars spanning a large Galactic disk radius. The sample used here does not include many stars in the inner disk, which may introduce a bias in our derivation of the central metallicity evolution. 
Future improvement on age and metallicity measurements will eventually provide better calibration of Eq.~\ref{eq3} and the central metallicity evolution. 

\section{Conclusions}
\label{subsec:future}

In this work we showed that stellar birth radii can be derived directly from the data with minimum prior assumptions on the Galactic enrichment history. We first developed an empirical method to recover the time evolution of the MW metallicity gradient, $\mathrm{\nabla [Fe/H](r,\tau)}$, through its inverse relation to the metallicity range as a function of age today, using a high-precision large data set of MW disk subgiant stars from the LAMOST survey. Using the age and metallicity measurements in our sample, we could place stars back to their birth radius, \rbir. Our main findings can be summarized as follows:

\begin{itemize}
    \item We found a steepening of the birth metallicity gradient from 11 to 8 Gyr ago (Fig.~\ref{fig:fehevo}), which coincides with the time of the MW last massive merger, GSE.

    \item The decrease in metallicity with age of the AMR upper boundary (Fig. \ref{fig:amr}), which can be misinterpreted as a chemical dilution in the last 8 Gyr, is a result of the superposition of different \rbir\ tracks, reversing the well-defined negative slopes of the underlying mono-\rbir\ AMRs. This is an example of Simpson’s paradox -- a commonly encountered statistical phenomenon in the field of Galactic Archaeology \citep{Minchev2019}. 

    \item We found a faster decrease in \alphafe\ for the innermost \rbir\ bins ($\sim6$ kpc, see Fig.~\ref{fig:afefeh}a), which can be explained if star formation in the innermost few kpc started earlier, thus reaching sooner the Type Ia Supernova dominated evolutionary phase that causes the fast drop in \alphafe. This suggests that inside-out formation was already taking place within this very concentrated disk and very early on (about 5 kpc in radius and 11 Gyr ago).

    \item The most metal-poor stars in the \lowalpha\ disk (at [Fe/H] $\lesssim-0.4$ dex) formed in the outer disk, during or after the GSE merger, indicating that these stars could be formed directly from the gas brought in by the merger \citep[also see][]{Buck2023}. 
    
    \item The gap between the high- and the low-\alphafe\ sequences originates from the quick time transition between them. This may be associated with the GSE merger, as the time of crossing the gap for each \rbir\ track in Fig.~\ref{fig:afefeh}a coincides with the time period during which $\mathrm{\nabla [Fe/H]}$ steepens. 

    \item By dissecting the disk into mono-\rbir\ populations, we find clumps in the low-\alphafe\ sequence (Fig.~\ref{fig:afefeh}), which coincide in time with star-formation bursts associated with the Sagittarius dwarf galaxy \citep{RuizLara2020}.

    \item We estimated that the Sun was born at $4.5\pm 0.4$~kpc from the Galactic center.
    
\end{itemize}

Our data-driven technique not only recovers the trends in chemo-kinematic relations expected from detailed forward modeling \citep{matteucci89, Chiappini1997, Minchev2013, Sharma2021,Hemler2021, Lu2022, Buck2020, Buck2020a, Frankel2020}, but also paints a detailed picture of the MW disk formation, and in a completely different, simpler approach. Applying our method to larger datasets can help with calibration of chemical evolution models, by providing the detailed evolution with birth radius and cosmic time for many elements \citep[e.g.][]{Ratcliffe23b, Ratcliffe23c}.

A challenge for future work will be to account for all the physical processes that have contributed to the evolution of the metallicity as a function of radius and time that we recover in this work.

\section*{Acknowledgements}
The authors acknowledge Hans-Walter Rix, Maosheng Xiang, and Diane Feuillet for helpful comments. This work has used data products from the Guoshoujing Telescope (LAMOST). 
I.M. and B.R. acknowledge support by the Deutsche Forschungsgemeinschaft under the grant MI 2009/2-1. 
LAMOST is a National Major Scientiﬁc Project built by the Chinese Academy of Sciences. Funding for the project has been provided by the National Development and Reform Commission. LAMOST is operated and managed by the National Astronomical Observatories, Chinese Academy of Sciences. 
This work has made use of data products from the European Space Agency (ESA) space mission Gaia. Gaia data are being processed by the Gaia Data Processing and Analysis Consortium (DPAC). Funding for the DPAC is provided by national institutions, in particular the institutions participating in the Gaia MultiLateral Agreement. The Gaia mission website is https://www.cosmos.esa.int/gaia. The Gaia archive website is https://archives.esac.esa.int/gaia.
This publication has also used data products from the 2MASS, which is a joint project of the University of Massachusetts and the Infrared Processing and Analysis Center/California Institute of Technology, funded by the National Aeronautics and Space Administration and the National Science Foundation.
We gratefully acknowledge the Gauss Centre for Supercomputing e.V. (www.gauss-centre.eu) for funding this project by providing computing time on the GCS Supercomputer SuperMUC at Leibniz Supercomputing Centre (www.lrz.de).
This research was carried out on the High Performance Computing resources at New York University Abu Dhabi. We greatly appreciate the contributions of all these computing allocations. TB's contribution to this project was made possible by funding from the Carl Zeiss Foundation.
This research made use of Astropy,\footnote{http://www.astropy.org} a community-developed core Python package for Astronomy \citep{astropy:2013, astropy:2018}.

\vspace{5mm}
Facilities: Gaia, LAMOST


Softwares: Astropy \citep{astropy:2013, astropy:2018}, Numpy \citep{oliphant2006guide}, sklearn \citep{scikit-learn}.


\section*{Data Availability}
Data available on request.

\bibliographystyle{mnras}
\bibliography{references.bib} 




\appendix 
\label{sec:append}

\section{Testing our method on the NIHAO-UHD}
\label{sec:rb_test}

To test the robustness of our approach, we apply the method developed in Sec.~\ref{subsubsec:grad} to infer \rbir\ for the NIHAO-UHD g2.79e11 simulation. We already showed in Fig.~\ref{fig:1} that we could recover the radial birth gradient within 9\% for this model. We now use the \feh\ and age from the last snapshot to recover \rbir\ as we did for the data.

We first rescale the simulation to match the MW for better comparison, by multiplying all the positions ($x$, $y$, $z$) in the simulation by the ratio of the MW radial scale-length to the simulation scale-length, 3.5/5.6, and the all velocities ($v_x$, $v_y$, $v_z$) by ratio of the MW rotation curve to the that of the simulation, 240/340.  

The birth metallicity gradient is inferred using the metallicity range of mono-age populations, as we did for the MW and described in Sec.~\ref{subsubsec:grad}. Similar to the data, we use Eq.\ref{eq3} to derive the time evolution of the radial metallicity gradient by setting $b$ to be the metallicity gradient measured today (at the final simulation time) and adjusting the scaling factor, $a$, so that the maximum gradient matches up with that of the simulation. 
To infer the central metallicity evolution, we use the upper envelope of the AMR as described in Sec.~\ref{subsubsec:grad} (using 95\% [Fe/H] of the central region, same selection as in the data).

\begin{figure}
    \centering
    \includegraphics[width=0.48\textwidth]{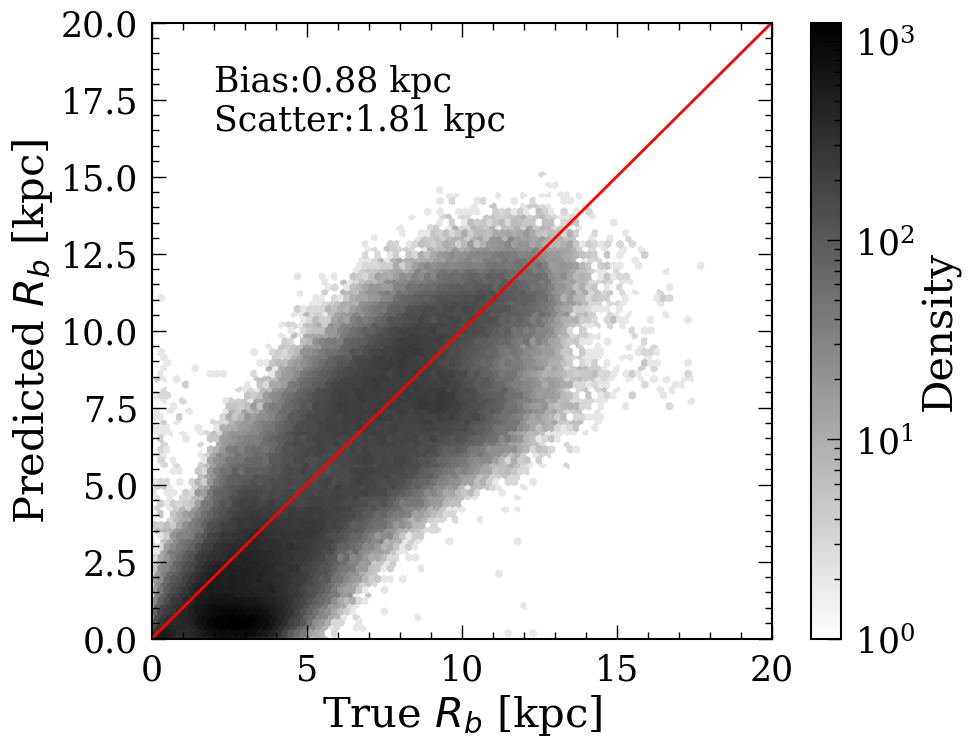}
    \caption{Comparing true \rbir\ with inferred \rbir\ for the NIHAO-UHD g2.79e12 simulation. We can recover \rbir\ within $\sim$ 20\% for stars in the solar neighborhood and beyond. }
    \label{fig:A5}
\end{figure}

Applying the same method we applied to the data, we were able to infer birth radii for this galaxy just from the last snapshot, for stars born once a rotationally supported disk has started to form $\sim$10 Gyr ago \citep{Buck2020}, with an average scatter of 1.84 kpc and a bias of 0.24 kpc (Fig.~\ref{fig:A5}). 

It is worth pointing out that the uncertainty is almost constant across all radii, meaning the percentage uncertainty is smaller for stars that are born in the outer disk. In this particular simulation, we are able to recover \rbir\ within 20\% uncertainty for stars in the solar neighborhood and beyond. 

\begin{figure*}
    \centering
    \includegraphics[width=0.98\textwidth]{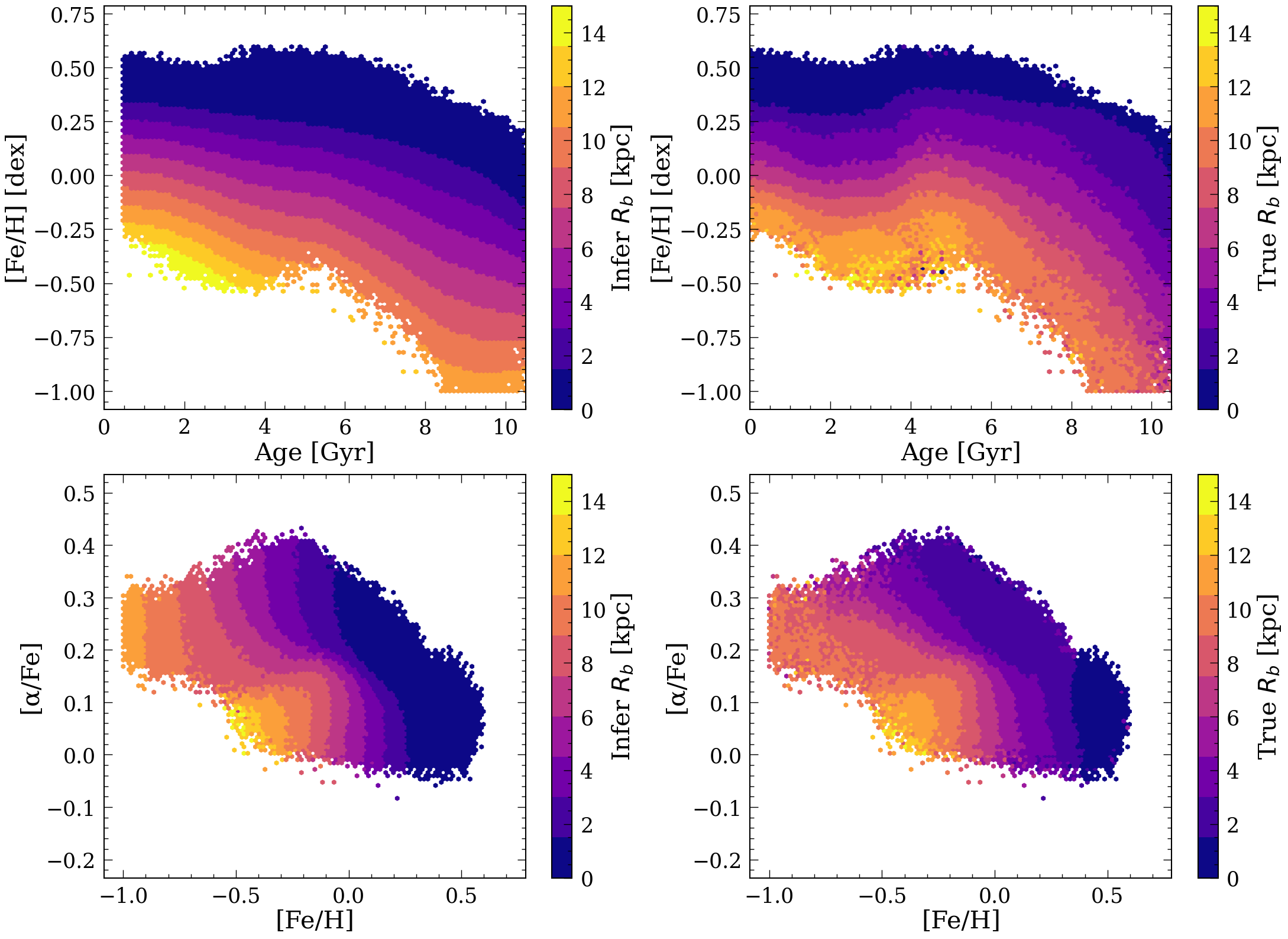}
    \caption{Age-metallicity-relation (AMR; top row) and \alphafe-\feh\ plane (bottom row) colored by inferred \rbir\ (left column) and true \rbir\ (right column).
    We are able to recover the general trend of the AMR and \alphafe-\feh\ in different \rbir.
    For example, the smaller range in \rbir\ in the AMR for stars born between 8-10 Gyr ago compared to that for a more recent time due to the steepening in the metallicity gradient;
    the slight dip in the AMR 3-4 Gyr ago;
    the impressive resembling of the evolution in mono-\rbir\ populations, especially for stars born in the outer disk, in the \alphafe-\feh\ plane. 
    It is worth noting that a merger plunged into the disk for this simulation $\sim$ 5 Gyr ago, which disrupted the linear relation between the metallicity gradient at that lookback time and the Range[Fe/H] at the same mono-age population (see Fig.~\ref{fig:1} top left plot).
    Since this most likely did not happen in the MW, we do not expect a deviation as larger as that seen in this simulation.}
    \label{fig:A6}
\end{figure*}

In Fig.~\ref{fig:A6}, we show the AMR (top row) and \alphafe-\feh\ plane (bottom row) colored by inferred \rbir\ (left column) and true \rbir\ (right column). The major features in the AMR and \alphafe-\feh\ plane are nicely captured by the inferred \rbir.
For example, the dilution of the metallicity in mono-\rbir\ populations shown in the AMR at $\sim$ 3-4 Gyr, and the enhancement of \alphafe\ around the transition between high- to \lowalpha\ (\alphafe\ $\sim$ 0.15) for stars in mono-\rbir\ populations shown in the \alphafe-\feh\ plane.

This test provides strong supporting evidence that our method is robust, especially for stars born in the outer disk. Given that the simulated disks are in general kinematically hotter, the intrinsic scatter around the metallicity gradient is expected to be larger compared to that of the MW \citep{Lu2022}. As a result, using this method to infer birth radii in the MW disk should yield an even smaller uncertainty. 

\bsp	
\label{lastpage}
\end{document}